\documentclass[]{aastex701}
\usepackage{natbib, twoopt}
\usepackage{float}
\usepackage{graphicx}
\usepackage{epstopdf}
\usepackage{enumitem}
\usepackage[caption=false]{subfig}
\usepackage{xcolor}
\usepackage{gensymb}

\newcommand{\revfont}{
  \bfseries 
  \color{red}
}
\DeclareTextFontCommand{\rev}{\revfont}
\graphicspath{{Figures/}}
\RequirePackage{import}
\usepackage{lineno}

\usepackage{textcomp}
\usepackage{gensymb}

\newcommand{\yxfont}{
\color{blue}
\bfseries 
}

\DeclareTextFontCommand{\yx}{\yxfont}

\begin{document}

\title{Collective Characterization of Polar Crown Filament Eruptions}
 
    \author{Ying Wang}
	\affil{Institute for Space Weather Sciences, New Jersey Institute of Technology, 323 Martin Luther King Boulevard, Newark, NJ 07102-1982}
    \email{yw26@njit.edu}

    \author{Yan Xu}
	\affil{Institute for Space Weather Sciences, New Jersey Institute of Technology, 323 Martin Luther King Boulevard, Newark, NJ 07102-1982}
	\email{yan.xu@njit.edu}
    
	\author{Ju Jing}
	\affil{Institute for Space Weather Sciences, New Jersey Institute of Technology, 323 Martin Luther King Boulevard, Newark, NJ 07102-1982}
	\email{ju.jing@njit.edu}

	\author{Vasyl Yurchyshyn}
	\affil{Big Bear Solar Observatory, New Jersey Institute of Technology, 40386 North Shore Lane, Big Bear City, CA 92314-9672, USA}
	\email{vasyl.yurchyshyn@njit.edu}
    
	\author{Haimin Wang}
	\affil{Institute for Space Weather Sciences, New Jersey Institute of Technology, 323 Martin Luther King Boulevard, Newark, NJ 07102-1982}
	\affil{Big Bear Solar Observatory, New Jersey Institute of Technology, 40386 North Shore Lane, Big Bear City, CA 92314-9672, USA}
    \email{haimin.wang@njit.edu}
    
\begin{abstract}

Polar crown filaments (PCFs) are crucially related to the structure of solar polar magnetic fields. They are physically significant because of the ``rush-to-the-pole'' motion, which is an indicator of the poleward drift of magnetic fields from decayed active regions. The formation and evolution of PCFs offer important insights into the Sun's magnetic activity at high latitudes and the solar cycle variation. In this study, we present a collective analysis of 50 PCF disappearance events covering solar cycle 24 and 25. Among them, 24 show rotational motions, and nine have sufficiently clear magnetic configurations for comparison with numerical model predictions. In all nine events, the observed rotation directions agree with the model predictions, suggesting possible magnetic-flux-rope-like structures in these events.

All except one of the PCFs in our sample disappeared through eruption rather than thermal disappearance, and most associated CMEs were slow ($<800$ km s$^{-1}$). For events accompanied by transient coronal holes, the estimated magnetic flux associated with the dimming regions is on the order of $10^{20}$--$10^{21}$ Mx. Time--distance analysis identifies 22 events with a clear slow-rise to fast-rise transition, with onset heights ranging from several tens to about 160 Mm and a mean onset velocity of about 2.8 km s$^{-1}$ after excluding the highest-velocity event. Together, these findings characterize the magnetic structure and dynamics of PCF eruptions.

\end{abstract}



\section{Introduction}

\label{Sect:intro}

Filaments are vast, dark structures of cool, dense plasma suspended above the solar surface and supported by magnetic fields. Polar crown filaments (PCFs) are characterized as a special group of filaments. They reside near the polar areas, at high latitudes, forming a ring-like structure around the poles \citep{Hansen1975a}. Filaments usually locate right above polarity inversion lines (PILs). The PILs under PCFs are distinctive because they separate the polar magnetic fields (predominant) from the decayed sunspot group drifting away from the equator. Therefore, PCFs are a good proxy of high latitude PILs on a large scale. Statistical studies have shown a distinct horizontal orientation of PCFs which is then used as key factor to identify PCFs \citep{Hyder1965}. The PCF distribution in the northern and southern hemispheres are not even \citep{Waldmeier1971, Hansen1975} and a temporal offset was noticed between PCFs and polar magnetic fields \citep{Chatterjee2017, Gopalswamy2018, Xu2018b}. More importantly, the poleward migration of PCFs, well known as ``rush to the pole'' \citep{Lockyer1931, Hyder1965},  is a signature of interaction between the old-cycle polar fields and the new-cycle diffused active region fields in the process of polar field reversal \citep{Leroy1983,Tang1987,Panesar2014}. To understand the consequence of polar field shrink, the poleward migration of PCFs has been studied extensively \citep{Ricco1892, Ananthakrishnan1952, Fujimori1984, Lorenc2003, Cliver2014, 
Gopalswamy2018, Xu2018b}.

Unlike the temporal and spatial distribution of PCFs, the eruptions were less studied. Filament eruptions can release immense amounts of energy and matter into space and are significant drivers of coronal mass ejections (CMEs), which are large expulsions of plasma and magnetic field from the Sun's corona. In the literature, eruptions from active region filaments usually accompanied with flares and attracted significant attention. As they are far away from solar active regions, PCF eruptions are classified as quiet Sun eruptions \citep{Martin1998, Diercke2019}. Due to the high latitude, the CMEs caused by the PCF eruption are less likely to reach Earth, making these events less attractive. On the other hand, PCFs usually have limb segments that help understand the vertical configuration of filaments. During an eruption, the side-view also reveals the connection of prominence material, CME cavity and the dynamics \citep{Regnier2011}. Moreover, PCF eruptions were believed to have impact on the solar polar reversal \citep{Gopalswamy2003a}. It was argued by these authors that filament eruptions would remove closed fields, and consequently open fields could be pushed to the pole for the magnetic polarity reversal. Studying PCF eruptions thus provides valuable insights into the Sun's global magnetic field and its long-term evolution. 

\section{Data Mining}

Filaments appear as obvious dark features when observed in the H$\alpha$ line at 6563~\AA\ and EUV 304~\AA\ line. The formation temperatures are 10,000 K for H$\alpha$ line and about 80,000 K for EUV 304~\AA  \citep{Wang1998}. Because of the higher temperature, EUV 304~\AA\ line observes chromospheric features in higher altitude \citep{Wang1998, Xu2010a}. Automatic or machine learning based methods have been developed to identify filament and their eruptions  \citep{Qu2005, Hao2016, Chatterjee2017, Diercke2019, Chatterjee2020}. However, PCFs show relatively low contrast in EUV 304~\AA\ and highly segmented in H$\alpha$ line (possibly due to the lower height), making the automatic detection less efficient. Therefore, we select the visual method to find the PCF eruptions by checking SDO/AIA 304 daily movies. This method utilizes the human detection to compromises the low contrasts in EUV 304~\AA\ and takes the advantage of SDO/AIA's long duty cycle. We followed the criteria described in \citet{Xu2018b} to detect PCFs. Specifically, 1) a latitude threshold of 50 degree is used to exclude  low-latitude filaments, which is a compromise among the different thresholds used in previous studies  \citep{Babcock1955, Brajsa1991, Hao2015, McCauley2015}; 2) filaments near active regions or plage areas are excluded; 3) filaments with tilt angle greater than 25 degrees are excluded \citep{Leroy1983, Tlatov2016, Xu2021a}. Using the above criteria, 50 PCF disappearance events were identified during the observation period. Among them, 49 events showed clear eruption signatures, indicating that approximately 98\% of the PCFs disappeared through eruption. Table~\ref{list} lists the 50  PCF disappearance events.

\begin{figure}[ht]

\includegraphics{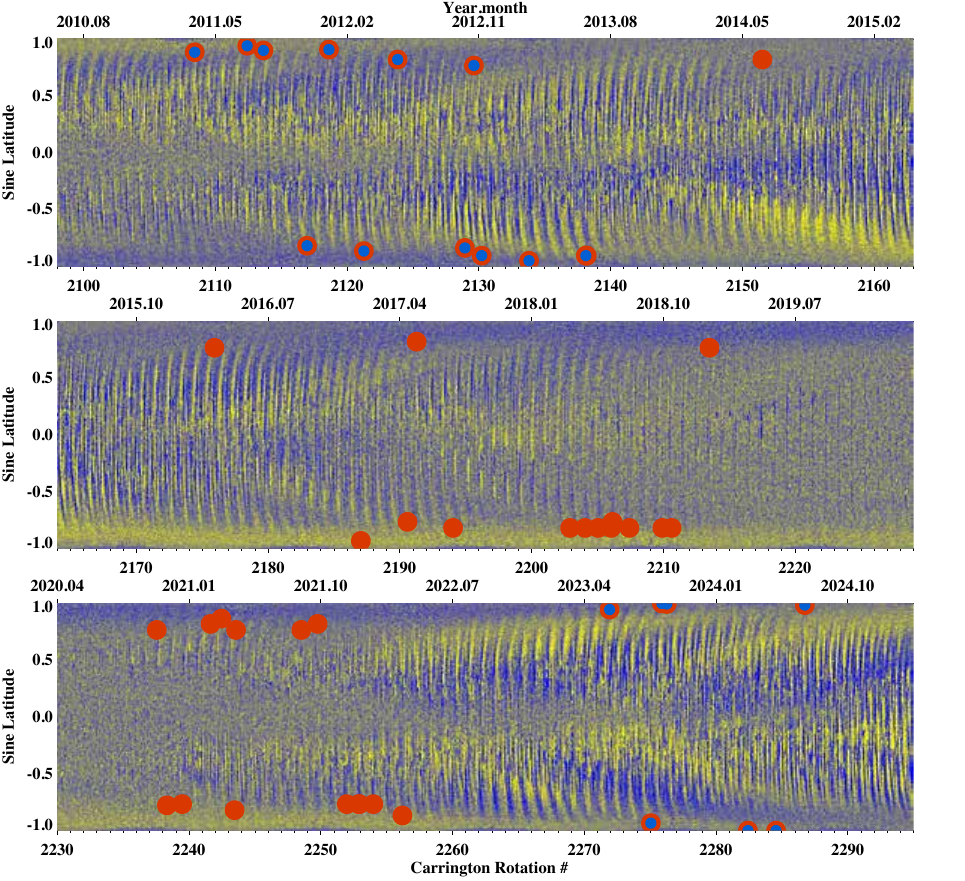}
\caption{\textsl{PCF eruptions superimposed on supersynoptic maps. Red filled dots: PCF eruptions occurring between polar field reversal periods, characterized by stable PIL latitudes. Red dots with blue centers: PCF eruptions occurring prior to polar field reversal, characterized by poleward migration of the PILs. The yellow color indicates negative magnetic polarity and the blue represents positive magnetic polarity.}}
\label{Fig:butterfly}
\end{figure}
\begin{table}[ht]
\begin{center}
\caption{PCF Disappearance Event List. \label{list}}
\label{eventlist2}

\begin{tabular}{lccccc}
\hline\noalign{\smallskip}
\textbf{Date \& Time} & \textbf{Latitude} & \textbf{CME Speed} & \textbf{Filament Rotation} & \textbf{TCH}  \\
\textbf{[UT]} & \textbf{[$^\circ$]} & \textbf{[km s$^{-1}$]} & [AIA 304~\AA]  &  [AIA 211~\AA] \\
\hline\noalign{\smallskip}

2011-Mar-27 19:52 & N62 & 191 & Rotated & Uncertain \\
2011-Jul-14 10:30 & N70 & 191 & Rotated & No \\
2011-Aug-17 03:00 & N64 & 692 & Not clear & No \\
2011-Nov-15 12:24 & S55 & --- & Not clear & Uncertain \\
2011-Dec-30 18:30 & N65 & 960 & Not clear & Uncertain \\
2012-Mar-12 00:59 & S60 & 638 & Not clear & Yes \\
2012-May-21 09:00 & S55 & 645 & Not clear & Uncertain \\
2012-Oct-07 05:30 & S57 & 663 & Rotated  & Yes \\
2012-Oct-26 09:00 & S50 & --- & Not clear & Uncertain \\
$^{*}$2012-Nov-01 22:30 & S60 & --- &  Not clear& --- \\
2012-Nov-10 22:30 & S65 & 690 & Not clear & Uncertain \\
2013-Feb-16 20:30 & S72 & 150 & Rotated & Uncertain \\
2013-Jun-14 16:00 & S65 & 228 & Rotated & Uncertain \\
2013-Jun-15 23:30 & S65 & 162 & Not clear & Uncertain \\
2013-Jun-16 06:00 & S65 & 152 & Not clear & Uncertain \\
2014-Jun-15 00:39 & N55 & 334 & Not clear & Yes \\
2016-Apr-10 23:50 & N50 & 291 & Rotated & Yes \\
2017-Feb-08 00:30 & S69 & 104 & Rotated & No \\
2017-May-15 13:13 & S50 & 80 & Rotated & Yes \\
2017-Jun-03 10:30 & N55 & 227 & Rotated & No \\
2017-Aug-17 20:30 & S55 & --- & Not clear & No \\
2018-Apr-17 04:00 & S55 & 250 & Rotated & No \\
2018-May-18 03:13 & S55 & 250 & Rotated & Uncertain \\
2018-Jun-14 10:00 & S55 & 370 & Not clear & No \\
2018-Jul-11 07:00 & S55 & --- & Rotated & Uncertain \\
2018-Jul-13 23:50 & S50 & 232 & Not clear & Uncertain \\
2018-Aug-18 01:00 & S55 & 448 & Not clear & No \\
2018-Oct-25 06:00 & S55 & 169 & Rotated & Uncertain \\
2018-Nov-13 12:00 & S55 & 405 & Not clear & No \\
2019-Jan-30 17:00 & S50 & --- & Rotated & Yes \\
2020-Nov-16 13:30 & N50 & 220 & Rotated & No \\
2021-Jan-08 04:00 & S50 & 393 & Rotated & Yes \\
2021-Mar-08 00:00 & N55 & --- & Rotated & No \\
2021-Mar-30 13:00 & N60 & --- & Rotated & Yes \\
2021-Apr-26 15:30 & S55 & 120 & Rotated & Yes \\
2021-Apr-29 16:18 & N50 & 88 & Rotated & No \\
2021-Sep-12 12:24 & N50 & 485 & Rotated & No \\
2021-Oct-15 16:00 & N55 & 251 & Not clear & Yes \\
2021-Dec-15 00:55 & S50 & 216 & Rotated & No \\
2022-Jan-09 08:52 & S50 & 215 & No data & No data \\
2022-Jan-10 12:12 & S50 & 249 & Not clear & Uncertain \\
2022-Feb-08 14:36 & S50 & 182 & Not clear & Uncertain \\
2022-Apr-09 06:30 & S60 & 163 & Not clear & No \\
2023-Jun-12 03:18 & N71 & 500 & Not clear & Uncertain \\
2023-Sep-05 15:00 & S69 & 144 & Rotated & Uncertain \\
2023-Sep-28 09:00 & N88 & 205 & Rotated & No \\
2023-Oct-08 13:00 & N84 & 320 & Not clear & No \\
2024-Mar-24 14:30 & S88 & 306 & Not clear & No \\
2024-May-21 17:30 & S88 & 397 & Not clear & No \\
2024-Jul-20 10:00 & N80 & --- & Not clear & No \\
\hline\noalign{\smallskip}
\end{tabular}
\par\smallskip
{\footnotesize $^{*}$Disappeared through thermal disappearance}
\end{center}

\end{table}

\section{Results}

\subsection{General Observations}

Filament eruptions are usually linked with CMEs. They are among the most important solar eruptive events that lead to space weather effects. \citet{Jing2004} studied 98 filament eruptions and found that 55\% of them were associated with CMEs. More specifically, quiet Sun (QS) filaments had a higher correlation (54\%) with CMEs, while 43\% of active region (AR) filaments led to CMEs. \citet{Gilbert2000} showed a similar correlation (46\%) between AR filaments and CMEs. These results indicate that the QS filament eruptions tend to have a better chance evolving to CMEs. A straight forward understanding is that lack of strong strapping fields above flux ropes make QS filaments easier to run away from the solar surface. 

For each event in Table~\ref{list}, we checked the corresponding CME in SOHO/LASCO CME list (\url{https://cdaw.gsfc.nasa.gov/CME_list/}), by comparing the occurrence time and the position angle. In our list, we found that 41 PCF eruptions were associated with identifiable CMEs. This provides a more direct inference, as PCFs are a special type of QS filaments located in areas with much weaker magnetic fields. According to \citet{Zou2019}, CMEs with speed greater than 800 km s$^{-1}$ are classified as fast, whereas those with speed below 800 km s$^{-1}$ are considered slow. Fast CMEs are often associated with active-region eruptions and major flares, where strong magnetic reconnection can contribute to rapid acceleration. In contrast, slow CMEs are frequently associated with eruptive prominences and quiescent filament eruptions \citep{Gosling1976}. Because our selection excludes filaments near active regions, the associated CMEs are expected to be predominantly slow events. Consistent with this expectation, most CMEs associated with the selected PCF eruptions are slow, with an average speed of about 316 km s$^{-1}$.Another mechanism that can cause filaments disappear is thermal disappearance \citep{Mouradian1995}, by which the filament plasma is heated to a certain level so that the filaments become less obvious or vanish in certain wavelengths. When the heating ceases and the plasma cools down, the filaments may reappear. Only one of the 50 PCF disappearance events was classified as thermal disappearance, while the remaining events were associated with eruptions. This dominance of eruptive events may be related to the relatively weak magnetic fields in the polar regions, where the weaker overlying coronal field provides less confinement, allowing PCFs to erupt more easily.

In Figure~\ref{Fig:butterfly}, the PCF eruption events are superimposed on a supersynoptic diagram of line-of-sight magnetograms. This supersynoptic diagram composes 198 HMI synoptic maps, spanning Carrington Rotations 2098 to 2295 (covering May 2010 to end of 2024), and illustrates the evolution of the Sun's magnetic field over time \citep{Svalgaard_2013}. The blue and yellow colors indicate positive and negative polarities, respectively. It is evident that the PCF events occurred near PILs, which mark the boundaries separating the polar magnetic fields from the decayed magnetic fields migrating from lower latitudes. Two distinct symbols are used to classify the PCF eruptions. Red filled circles indicate events that took place during the mid-point between two polar reversals, while red circles with a blue center represent events closer to the time of a polar magnetic reversal. During the period of a polar magnetic reversal, the polar PIL was observed to move distinctly toward higher latitudes. Figure~\ref{Fig:butterfly} covers two polar reversals in solar cycle 24 and 25. The first reversal, which marked the end of Solar Cycle 24's magnetic polarity, was characterized by significant asymmetry between the hemispheres. The magnetic field at the north pole experienced a double reversal between 2012 and 2014 \citep{Shiota2012, Karna2014, Sun2015}. The magnetic field at the South Pole reversed much later, in 2015 \citep{Sun2015, Gopalswamy2016}. Several PCF eruptions were observed near the North Pole prior to the initial reversal. However, no PCF eruptions were observed near the North Pole close to the second (subsequent) reversal event. In our sample, PCF eruptions were observed both shortly before polar reversals and during relatively stable periods between reversals. However, the limited number of events and reversals covered does not allow us to establish a general relationship between PCF eruptions and polar reversals.

\subsection{Rotation of PCF Eruptions}

Rotational motions during filament and flux-rope eruptions have been widely reported in previous studies and are considered an important manifestation of the conversion of magnetic twist into writhe \citep[e.g.,][]{Green_2007, Romano_2003, Dhara_2014}. Numerical modeling further suggests that the rotation sense is closely related to the magnetic helicity sign and to the asymmetry of the coronal background field \citep[e.g.,][]{Lynch2009}. In general, left-handed flux ropes tend to rotate counterclockwise during the eruption, whereas right-handed structures typically rotate clockwise. These factors provide a useful diagnostic linking the observed rotation to the underlying magnetic topology.

\begin{figure}[ht]
\centering
\includegraphics[width=0.9\textwidth]{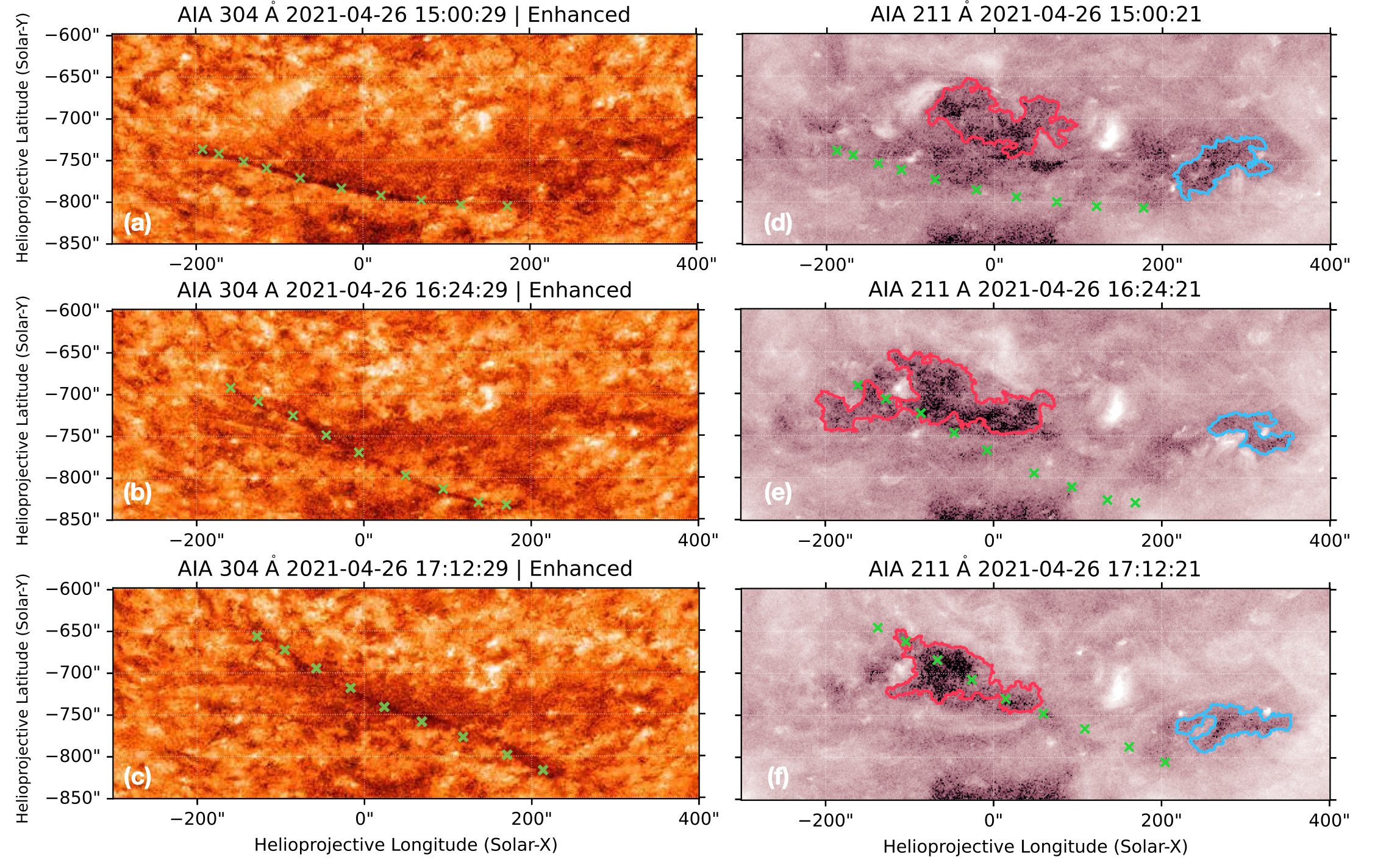}
\caption{AIA observations of the 2021 April 26 PCF eruption.
(a)--(c) AIA 304~\AA\ images showing the filament evolution and its clockwise rotation.
(d)--(f) AIA 211~\AA\ images with the tracked filament position (marked by ``$\times$'') and the positive (red) and negative (blue) dimming footpoints. 
The orientation angle of the filament changes from \(\sim166^\circ\) to \(\sim153^\circ\), corresponding to an increase in the acute tilt angle relative to the local east--west direction from \(\sim14^\circ\) to \(\sim27^\circ\), consistent with right-handed magnetic helicity.}

\label{Fig:comparison}
\end{figure}
In our sample of 49 PCF eruptions (Table~\ref{list}), 24 events appear to show rotational motion during eruption. However, the relatively low contrast in AIA images and the complex motion of the erupting filaments make it difficult to determine the rotation direction in all events. To determine the rotation direction, we manually identified the leading edge of the filament in successive contrast-enhanced AIA images and tracked its motion across consecutive frames. Among the 24 events showing rotational motion, 16 exhibited sufficiently clear and coherent rotation to distinguish between clockwise and counterclockwise directions as viewed from SDO. No projection correction was applied; the analysis focused on the apparent rotation direction rather than the three-dimensional rotation angle. In some events, the rotation is clear, but the dimming signatures are absent or too weak to be distinguished from the background. Ultimately, we identify nine events with clear rotation signatures for which the helicity sign can be inferred. These events include both cases with two well-defined transient coronal dimming footpoints and cases with only one identifiable dimming region, for which the conjugate footpoint is inferred from the surrounding magnetic polarities and the local PIL.

A representative two-footpoint example occurred on 2021 April 26 at high southern latitude. As shown in Figure~\ref{Fig:comparison}, panels~(a)--(c) display the filament evolution from 15:00 to 17:12~UT. To quantify the rotation, we manually tracked the leading edge of the filament marked by ``$\times$''. Panels~(d)--(f) show the corresponding AIA 211~\AA\ images, where the filament position is marked by ``$\times$'', and the red and blue contours outline the positive- and negative-polarity dimming regions, respectively, which are interpreted as the conjugate footpoint regions of the erupting structure, following the similar approach described
by \citet{Wang2026}.. A clear clockwise rotation is evident throughout the eruption. From 15:00~UT to 17:12~UT, the filament tilt increased from approximately $14^\circ$ to $27^\circ$. Based on the footpoint locations and the background magnetic field, this event shows right-handed helicity, consistent with the MHD modeling results of \citet{Lynch2009}.

\begin{figure}[h]
\centering
\includegraphics[width=0.7\textwidth]{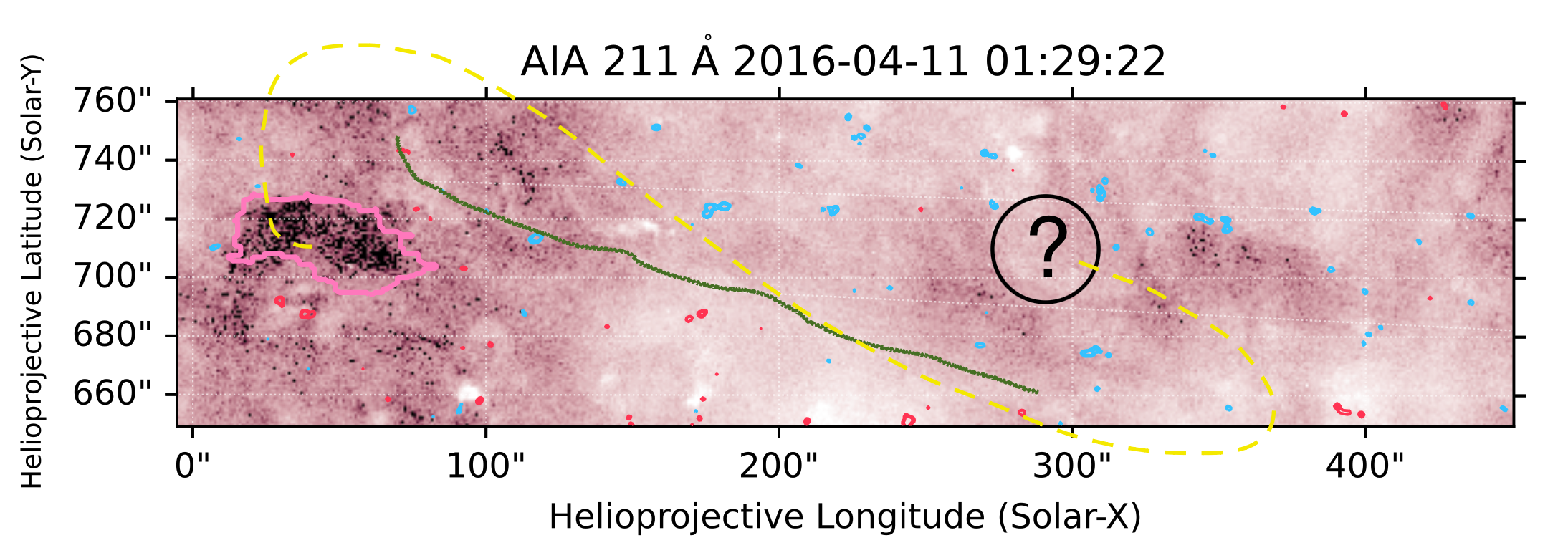}
\caption{Example showing how the conjugate footpoint can be inferred for an event with one visible dimming region. The red contour shows the positive-polarity dimming in the AIA image. The green line marks the PIL, and the yellow dashed line indicates the inferred structure.
}
\label{Fig:one_dimming}
\end{figure}

For events that showed clear rotation but only one identifiable dimming region, we inferred the helicity sign using the following method. As illustrated by the example event shown in Figure~\ref{Fig:one_dimming}, transient coronal dimmings are generally interpreted as signatures of the conjugate footpoints of the erupting magnetic structure. The background magnetic polarities are overplotted as blue and red dots, and the PIL is marked by the green line. One clear positive-polarity dimming could be identified in the AIA 211~\AA\ images.  We assumed that the conjugate footpoint of the observed positive-polarity dimming is rooted in the negative-polarity region on the opposite side of the local PIL. Under this assumption, the magnetic configuration is consistent with left-handed helicity. Following the same procedure, we inferred the helicity signs for the selected events with clear rotation signatures (see Table~\ref{rotation}).

The agreement between the observed rotation directions and the inferred helicity signs in these nine selected cases suggests that at least some PCF eruptions in our sample involve twisted, flux-rope-like magnetic structures. This interpretation is supported by magnetofrictional simulations \citep{Chen2024}, which show that sheared magnetic fields at high latitudes can produce magnetic flux ropes through reconnection.

\begin{table}[ht]

\centering
\footnotesize
\caption{Summary of Rotation, and Helicity Properties for the Selected PCF Events. \label{rotation}}
\begin{tabular}{lccccc}
\hline\noalign{\smallskip}
\textbf{Date \& Time[UT]} & \textbf{Latitude[$^\circ$]}   & \textbf{AIA Rotation}  & \textbf{Lynch's Result }& \textbf{Helicity }&   \\

\hline\noalign{\smallskip}
2011-Mar-27 19:52 & N62   & counter-clockwise           & Yes & L 
 \\

2012-Oct-07 05:30 & S57   & Clockwise           & Yes & R
\\ 

2016-Apr-10 23:50 & N50   & counter-clockwise           & Yes & L 
 \\
 
2017-May-15 13:13 & S50   & Clockwise           & Yes & R 
 \\

2019-Jan-30 17:00 & S50  & Clockwise          & Yes & R 
 \\

2021-Apr-26 15:30 & S55 & Clockwise    & Yes & R 
 \\

2021-Jan-08 04:00 & S50  & Clockwise   & Yes & R 
 \\

2021-Mar-30 13:00 & N60  & counter-clockwise & Yes & L 
\\

 2023-Sep-05 15:00 & S69  & Clockwise   & Yes & R 
 \\
\hline\noalign{\smallskip}
\end{tabular}
\end{table}

\subsection{Kinematic Measurements from Time--Distance Maps}
To further quantify the eruption kinematics, we constructed time--distance maps for the 49 erupted PCF events using AIA 304 \AA\ images. For each event, a linear slit was manually placed along the main propagation direction of the erupting filament. The intensity profile along the slit was extracted from each AIA image and then stacked in time to form a time--distance map. For each event, the slit started at the intersection of the solar limb and the line connecting the solar center to the pre-eruption apex. Its direction followed the apparent motion of the erupting structure in the EUV image sequence, and its length covered the region over which the structure could be reliably tracked. The resulting heights and velocities were measured in the image plane and converted to a projected physical distance assuming \(1^{\prime\prime}\approx725\) km. 

\begin{figure}[h]
\centering
\includegraphics[width=0.9\textwidth]{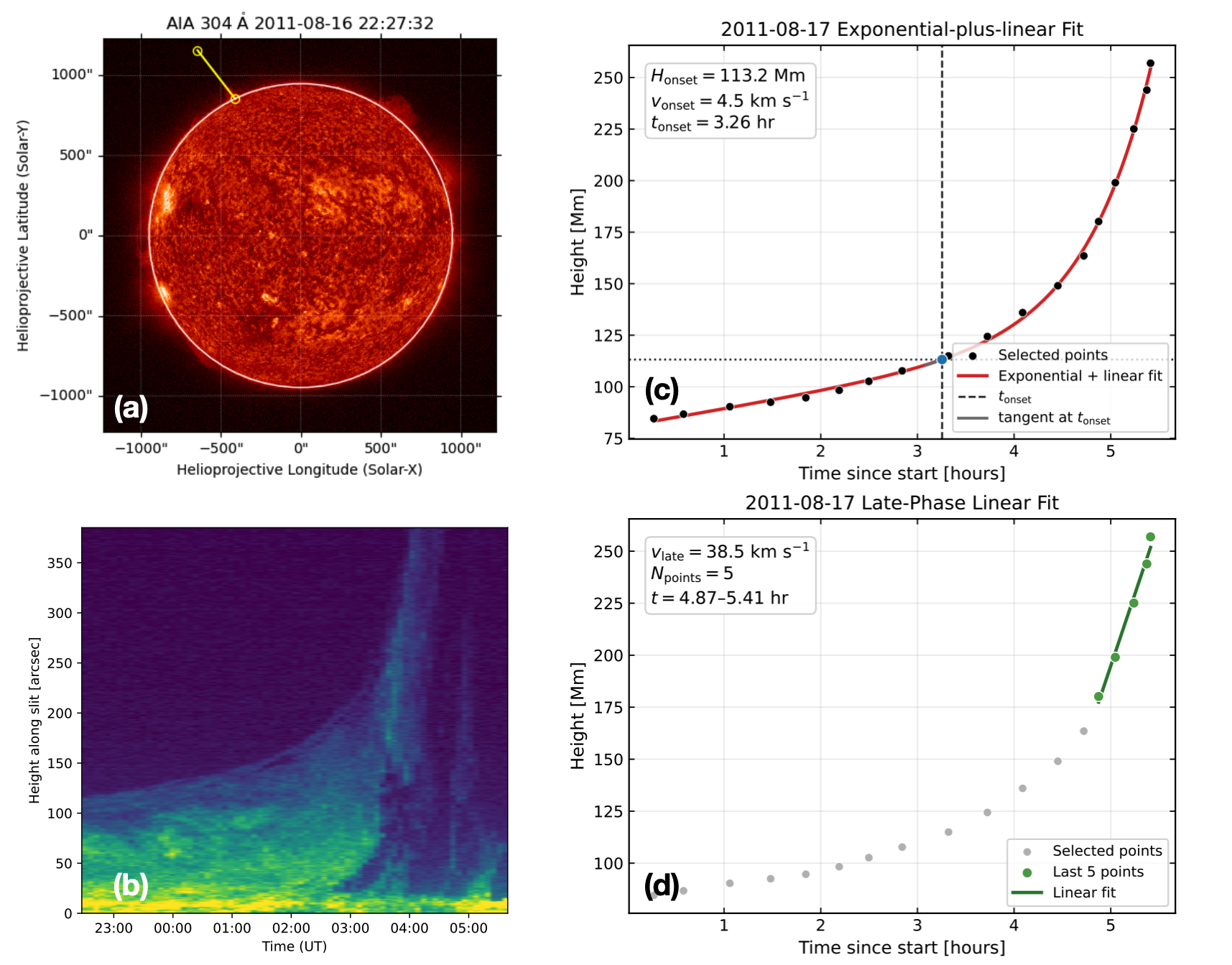}
\caption{
Example of the time--distance analysis for the 2011 August 17 event.
(a) AIA 304 \AA\ image with the manually selected slit.
(b) Time--distance map constructed by stacking the intensity profiles along the slit.
(c) Exponential-plus-linear fit to the selected height--time points. The vertical dashed line marks the fitted fast-rise onset time, and the horizontal dotted line marks the onset height.
(d) Linear fit to the last five selected height--time points, used to estimate the late-phase projected filament velocity.
}
\label{Fig:j-map}
\end{figure}

Figure~\ref{Fig:j-map} shows an example of the measurement procedure for the 2011 August 17 event. Panel (a) shows the AIA 304 \AA\ image with the selected slit. Panel (b) shows the corresponding time--distance map. The selected height--time points were fitted with an exponential-plus-linear function following \citet{Su2015},
\[
h(t)=c_0\exp\left(\frac{t-t_0}{\tau}\right)+c_1(t-t_0)+c_2 ,
\]
where \(h(t)\) is the projected height and \(t\) is time. This function provides a single empirical description of both the slow-rise and fast-rise phases. The velocity is obtained from the derivative of the fitted curve,
\[
v(t)=\frac{c_0}{\tau}\exp\left(\frac{t-t_0}{\tau}\right)+c_1 .
\]
The onset of the fast-rise phase is defined as the time when the exponential component of the velocity equals the linear component,
\[
t_{\rm onset}=t_0+\tau\ln\left(\frac{c_1\tau}{c_0}\right).
\]
The onset height and onset velocity were then evaluated as \(h(t_{\rm onset})\) and \(v(t_{\rm onset})\), respectively. For events in which the fitted parameters did not give a physically meaningful onset time, this onset measurement was not used in the onset statistics.

In addition to the onset velocity, we measured a late-phase projected filament velocity for comparison with the associated CME speed. This velocity was obtained by applying a linear fit to the last five manually selected height--time points for each event. This late-phase velocity represents an average plane-of-sky speed during the later part of the traced eruption and is therefore treated separately from the instantaneous onset velocity. An example of this linear fit is shown in Figure~\ref{Fig:j-map}(d).

\begin{figure}[h]
\centering
\includegraphics[width=1\textwidth]{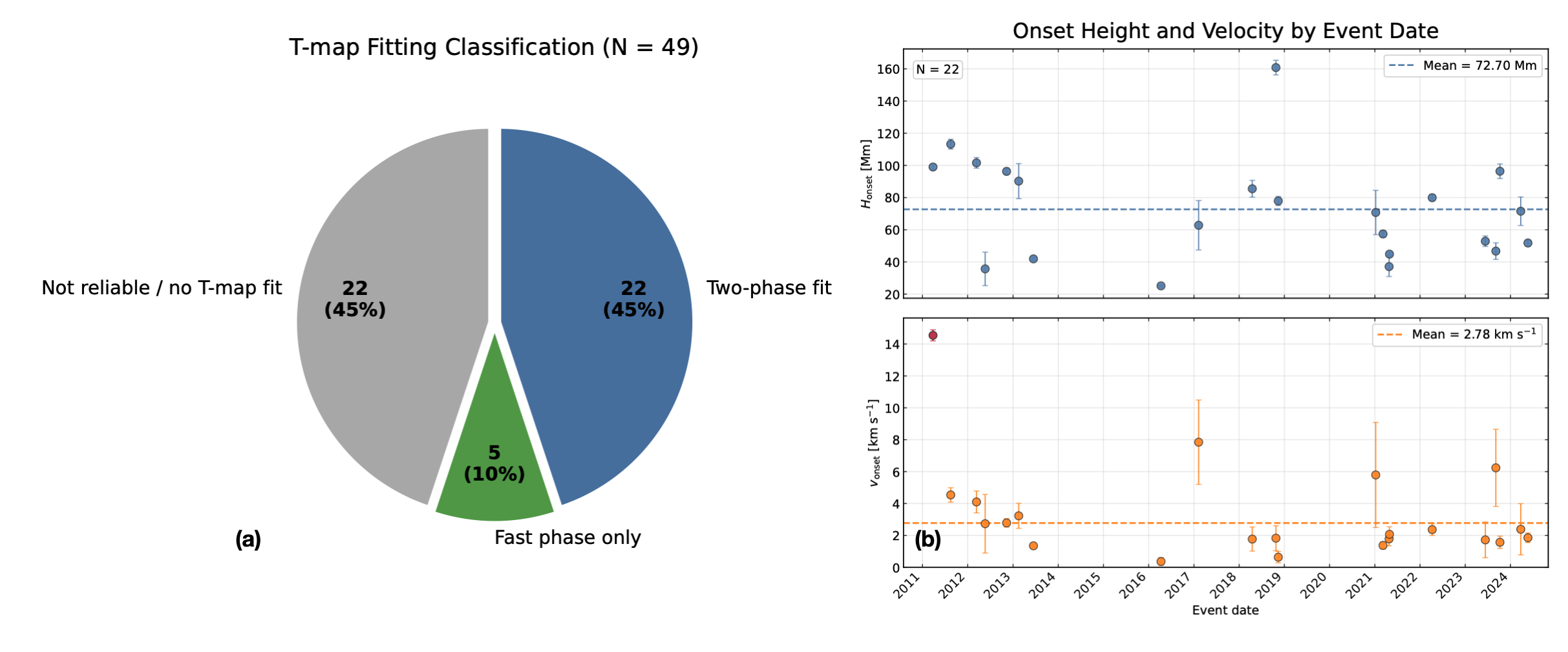}
\caption{Summary of the time--distance fitting results.
(a) Classification of the 49 PCF eruption events based on the quality of the time--distance fitting.
(b) Temporal distribution of the fitted onset height and onset velocity for the 22 PCF eruptions with reliable two-phase
time–distance fitting. (upper) Projected onset height Honset, measured along the manually selected slit as a function of event date.
(lower) Onset velocity vonset, obtained from the derivative of the exponential-plus-linear fit at tonset. Error bars represent the
uncertainties from the fitted height–time model. The dashed horizontal lines indicate the sample mean values; the
mean velocity is calculated after excluding the largest-velocity outlier.
}
\label{Fig:eventsummary}
\end{figure}

However, the two-stage eruption is not always obvious due to varying perspective angles and trajectory types. As Figure~\ref{Fig:eventsummary}(a) shows, among the 49 PCF eruption events, 22 events showed height--time profiles that were well described by the two-phase exponential-plus-linear fitting. Five events showed only a clear fast-rise phase and were characterized using the late-phase linear velocity only. For the remaining events, the fitting results were either not physically meaningful or the time--distance map did not provide a reliable height--time trajectory. For the 22 events with reliable two-phase fitting, the onset heights are mostly in the range of about 25--160 Mm, with a mean value of approximately 73 Mm. The onset velocities are generally small, with a median value of approximately 2.2 km s\(^{-1}\) and a mean value of approximately 2.8 km s\(^{-1}\), indicating that the filaments were still in the early stage of acceleration at the fitted onset time as shown in Figure~\ref{Fig:eventsummary}(b).

These values are consistent with previous statistical studies of prominence and filament eruptions observed by SDO/AIA. For example, \citet{McCauley2015} found an average fast-rise onset height of about 83 Mm and a slow-rise velocity of about 2.1 km s\(^{-1}\) for limb prominence and filament eruptions. Using stereoscopic data from STEREO, \citet{Xu2010a} demonstrated that a filament erupted at a speed of $12.2 \pm 1.2\text{ km s}^{-1}$ at an altitude of approximately 50 to 60 Mm. The comparable onset heights and low onset velocities found in our PCF sample suggest that PCF eruptions also commonly undergo a slow-rise to fast-rise transition. However, our sample is restricted to high-latitude PCF eruptions, which are closely related to the evolution of polar magnetic fields, and only a subset of the events show sufficiently clear height--time tracks for robust two-phase fitting.

We also compared both the onset velocity and the late-phase projected filament velocity with the associated CME speeds but no clear trend is found between the onset velocity and the CME speed in this limited sample. Similarly, the late-phase projected filament velocity measured does not show a significant correlation with the CME speed. The absence of a clear correlation suggests that the CME speed measured in the outer corona is not simply determined by the projected filament speed measured in the low corona. This may indicate that additional acceleration or deceleration can occur during the later CME evolution. The lack of correlation may also partly arise from projection effects and methodological differences between the filament and CME velocity measurements.

\subsection{Transient Coronal Holes associated with PCF Eruptions}

Transient coronal holes (TCHs) are sudden dimming in the corona associated with CMEs, which have been first detected in white light images several decades ago \citep{Hansen1974} and X-ray \citep{Rust1976} observations. Later on, similar dimming was noticed in EUV wavelengths, which are usually accompanied by EUV or EIT waves \citep{ Sterling1997,Thompson1998, Reinard2008, Mason2016}. The EUV observations and MHD modeling suggest that the TCHs are consequence of coronal material evacuation \citep{Zarro1999, Sterling2000, Cohen2009, Downs2012}. Therefore, TCHs are important in studying the mass, speed and energy of CMEs. Moreover, because the magnetic field lines are ``frozen'' into the plasma due to the plasma's high electrical conductivity and low plasma $\beta$,  TCHs are important for understanding the magnetic flux change associated with CMEs, especially near the poles where the measurements of magnetic field suffer significant uncertainties. \citet{Low1997} suggested that old magnetic flux in polar areas, which is residual from the previous solar cycle, can be removed by PCF eruptions. This hypothesis was supported by \cite{Gopalswamy2003a} findings. In this study, we estimate the level of magnetic flux removal associated with PCF eruptions and make a comparison with the polar magnetic flux. 

\begin{figure}[h]
\centering
\includegraphics[width=0.9\textwidth]{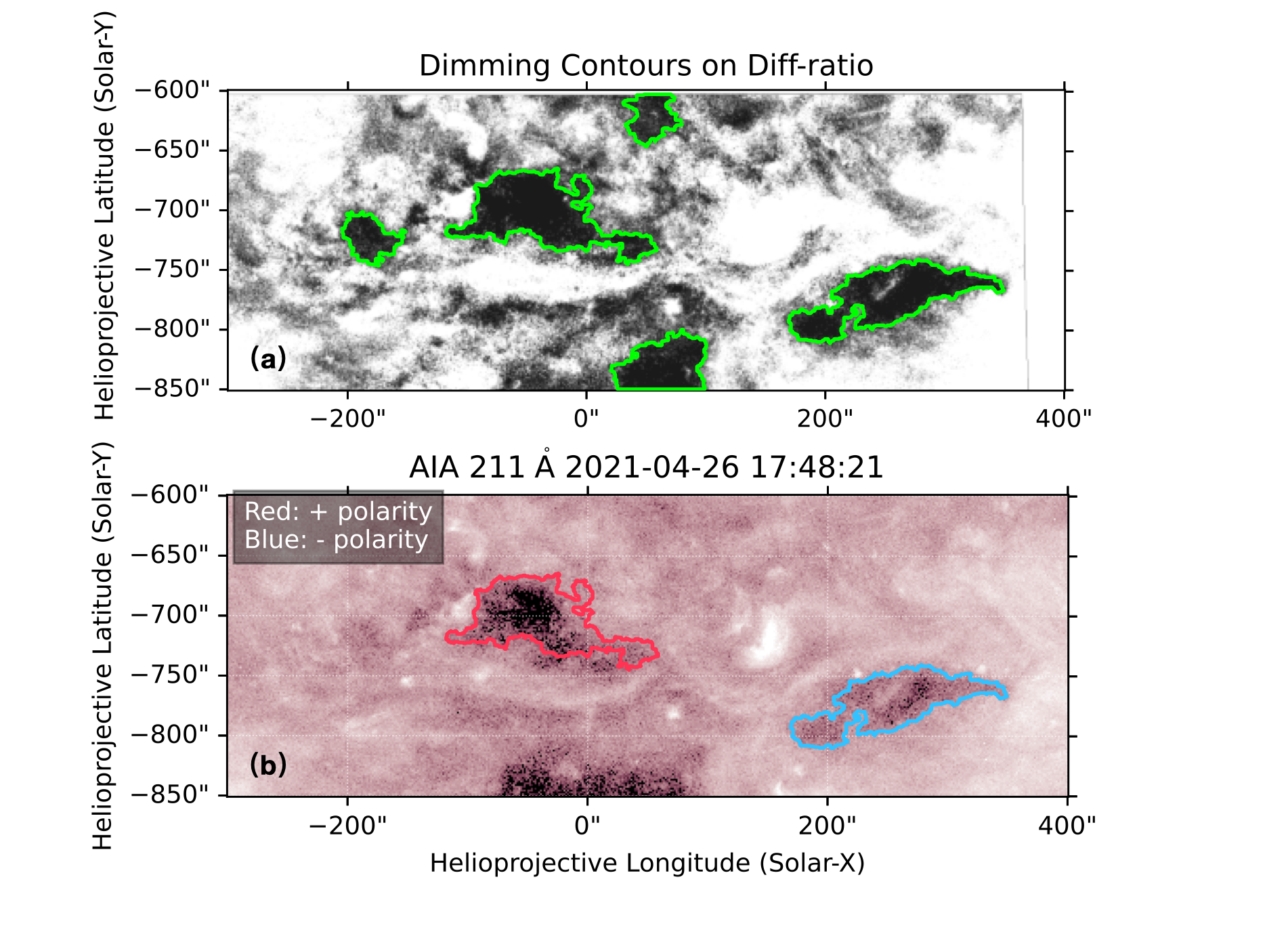}
\caption{
(a)  The 15\% dimmest contour of logarithmic base-ratio map relative to the pre-event background covered on the log-scaled base--ratio map.
(b) Dimming contours on the AIA 211~\AA\ image. Polarity-purity filtering using HMI leaves two single-polarity dimming regions, representing the true TCHs.
 }
\label{Fig:dimming}
\end{figure}

To identify the TCHs, we followed a standard approach that combines thresholding, difference imaging, and logarithmic scaling, as described in \citet{Thompson_1998}, \citet{Veronig_2006}, and \citet{Dissauer_2018}. First, we constructed a background pre-event AIA 211~\AA\ image by differentially rotating the pre-event frame to the event time. The pre-event image was selected at least three hours before the eruption and contained no strong active region within the target area. In case of PCF events, the background is typically a quiet-Sun region, which makes the difference image sensitive to both bright and dark structures in the base frame. Therefore, bright and dark outliers in each background map were clipped using the 2nd--98th percentile range to suppress transient artifacts.

The dimming strength was quantified using the base-ratio map relative to the background,
\[
R = \frac{I_{\rm target}}{I_{\rm base}},
\]
where \(I_{\rm target}\) and \(I_{\rm base}\) are the AIA 211~\AA\ intensities in the target and background images, respectively. A logarithmic scaling,
\[
L = \log_{10}(R),
\]

was then applied to the ratio map to enhance faint darkening signals and to identify dimming regions. Previous studies have commonly used a fixed logarithmic base-ratio threshold, such as \(L < -0.19\) \citep{Dissauer_2018}, which corresponds to an intensity decrease of about 35\%. However, PCF eruptions occur in quiet-Sun regions, where the pre-eruption EUV emission is relatively weak and the associated dimming signatures are often faint. In many of our events, the intensity decrease is smaller than 35\%, so a fixed threshold of \(L < -0.19\) would miss coherent dimming regions. Therefore, we define the TCH region using the 15\% dimmest pixels in the logarithmic base-ratio map. We also compared the dimming contours obtained from the logarithmic base-ratio and base-difference maps and found that, when applying a 10--15\% percentile threshold, the two methods produce very similar results.

Figure~\ref{Fig:dimming} shows an example of the TCH extraction. Panel~(a) displays the 15\% dimmest contour identified from the logarithmic base-ratio map. The selected dimming regions were further compared with the co-aligned HMI magnetogram to identify the positive- and negative-polarity footpoint regions. The TCH area was calculated from the number of pixels enclosed by the contour, with projection correction applied using the heliographic location of the region.

To estimate the uncertainty of the TCH area, we considered the sensitivity of the measured area to the selection of the target image. For each event, we obtained up to eleven AIA 211~\AA\ images within $\pm$5 min of the nominal target time. A logarithmic base-ratio map was constructed for each image using the same pre-event base image. The same dimming threshold, magnetic-polarity criterion, morphological filtering, and projection correction were then applied to all maps to determine the corresponding TCH areas.
The TCH area measured from the image at the nominal target time was adopted as the reported value. The uncertainty was estimated from the standard deviation of the N area measurements,
\[
\sigma_A =
\sqrt{
\frac{1}{N-1}
\sum_{i=1}^{N}
\left(A_i-\overline{A}\right)^2
},
\]
where \(A_i\) is the TCH area measured from the \(i\)-th target image, \(\overline{A}\) is the mean area of the N measurements, and \(N\) is the number of target images used for the uncertainty estimate. The resulting \(\sigma_A\) was adopted as the uncertainty of the measured TCH area. This uncertainty characterizes the combined effects of short-timescale area variability and the selection of the target image.

The mean magnetic field strength in the polar area is about 5--10 Gauss \citep{Benevolenskaya2004, Petrie2022}, and the distribution of magnetic fields is not uniform. Magnetic field strength in some small concentrations may reach several hundred Gauss to 1000 Gauss \citep{Tsuneta2008}. A typical strong-field patch has an angular size of about 3.5\arcsec, corresponding to $10^{15}$ to $10^{16}$ cm$^{2}$ \citep{Homann1997}, which is significantly smaller than the TCH area measured in this study. Therefore, we chose to use the mean field strength of 10 Gauss to estimate the magnetic flux in TCHs. The uncertainty of the magnetic flux was estimated by propagating the
uncertainties in both the TCH area and the magnetic field. Following
\citet{Liu_2012}, the random-noise upper bound of the HMI 720-s
line-of-sight magnetograms was taken to be \(\sigma_B=6.3\) G. To avoid
including pixels dominated by weak-field noise, \(N_p\) was defined as the
number of pixels within each TCH mask satisfying \(|B|\geq10\) G, consistent
with the magnetic-field threshold used for the polarity selection. The
uncertainty of the region-averaged magnetic field was then estimated as
\[
\sigma_{\bar{B}}=\frac{\sigma_B}{\sqrt{N_p}}.
\]
The total magnetic-flux uncertainty was calculated as
\[
\sigma_{\Phi}
=
\left[
(B\sigma_A)^2+
(A\sigma_{\bar{B}})^2
\right]^{1/2},
\]
where \(A\) is the TCH area, \(\sigma_A\) is the area uncertainty estimated
from the target-time variation, and \(B=10\) G is the characteristic magnetic
field used in the flux estimate.
To simplify the process, dimming pixels in each event listed in Table~3 were identified using a percentile-based threshold applied to the logarithmic base-ratio map. Small fragments and holes were removed using morphological operations. The HMI magnetogram was reprojected onto the AIA grid, and pixels with $|B| \ge 10$~G inside each dimming mask were used to compute a polarity purity for every connected component. Regions with significant single-polarity purity were retained as reliable single-polarity dimmings, representing the conjugate footpoints of the erupting flux rope.

As shown in Figure~\ref{Fig:dimming}(a), the initial dimming contour includes several disconnected regions, including a polar coronal hole. After evaluating the polarity purity of each candidate region and manually removing the polar coronal hole, only two dimming regions remained: the right region exhibits negative polarity and the left region exhibits positive polarity, corresponding to the two true TCH footpoints shown in Figure~\ref{Fig:dimming}(b).

To determine the physical area of each TCH, we projected the dimming mask into heliographic Stonyhurst (HGS) coordinate system, computed the mean longitude and latitude of each region, and applied geometric projection correction to convert the apparent pixel area into the estimated surface area on the Sun.

The results are listed in Table~\ref{TCH_full}. In the table, the $\Delta$ flux represents the difference between the positive and negative TCH magnetic fluxes; a positive $\Delta$ flux indicates that the flux associated with the $+$TCH is larger than that of the $-$TCH in our calculation. The hemisphere column shows the location of the filament, while the sign of the polar magnetic field indicates the polarity near the corresponding pole. However, clear pairs of dimming regions are difficult to identify near the limb and in the presence of nearby coronal holes. For the resulting sample of ten events, we find no clear relationship between the flux imbalance of the TCHs and the polarity of the nearby polar magnetic field. Given the limited sample size, this result should be interpreted with caution.
\begin{table}[ht]
\centering
\scriptsize
\setlength{\tabcolsep}{3.5pt}
\caption{TCH area and magnetic flux for selected events. \label{TCH_full}}
\label{TCH_full}

\begin{tabular}{lccccccc}
\hline\noalign{\smallskip}
\textbf{Date \& Time} &
\multicolumn{2}{c}{\textbf{$+$TCH}} &
\multicolumn{2}{c}{\textbf{$-$TCH}} &
\textbf{$\Delta$Flux} &
\textbf{Hem.} &
\textbf{PF} \\
\textbf{[UT]} &
\textbf{Area} &
\textbf{Flux} &
\textbf{Area} &
\textbf{Flux} &
\textbf{} & & \\
&
\textbf{[$10^{19}$ cm$^2$]} &
\textbf{[$10^{20}$ Mx]} &
\textbf{[$10^{19}$ cm$^2$]} &
\textbf{[$10^{20}$ Mx]} &
\textbf{[$10^{20}$ Mx]} & & \\
\hline\noalign{\smallskip}

2012-Mar-12 00:59 & $10.48\pm0.62$ & $10.48\pm0.66$ & $4.27\pm0.70$ & $4.27\pm0.71$ & $6.21\pm0.97$ & S & $+$ \\
2012-Oct-07 05:30 & $2.94\pm0.29$ & $2.94\pm0.30$ & $12.89\pm0.73$ & $12.89\pm0.74$ & $-9.94\pm0.80$ & S & $+$ \\
2014-Jun-15 00:39 & $3.02\pm0.07$ & $3.02\pm0.11$ & $1.45\pm0.07$ & $1.45\pm0.09$ & $1.57\pm0.15$ & N & $+$ \\
2016-Apr-10 23:50 & $1.18\pm0.21$ & $1.18\pm0.22$ & $1.67\pm0.27$ & $1.67\pm0.28$ & $-0.49\pm0.35$ & N & $+$ \\
2017-May-15 13:13 & $1.13\pm0.13$ & $1.13\pm0.14$ & $1.09\pm0.10$ & $1.09\pm0.11$ & $0.04\pm0.18$ & S & $-$ \\
2019-Jan-30 17:00 & $1.68\pm0.14$ & $1.68\pm0.16$ & $1.05\pm0.21$ & $1.05\pm0.22$ & $0.63\pm0.27$ & S & $-$ \\
2021-Apr-26 15:30 & $4.10\pm0.78$ & $4.10\pm0.78$ & $5.04\pm1.31$ & $5.04\pm1.31$ & $-0.94\pm1.53$ & S & $-$ \\
2021-Jan-08 04:00 & $7.82\pm0.23$ & $7.82\pm0.27$ & $10.18\pm0.31$ & $10.18\pm0.40$ & $-2.36\pm0.48$ & S & $-$ \\
2021-Mar-30 13:00 & $2.63\pm0.22$ & $2.63\pm0.24$ & $1.53\pm0.09$ & $1.53\pm0.11$ & $1.10\pm0.26$ & N & $+$ \\
2021-Oct-15 16:00 & $3.35\pm0.16$ & $3.35\pm0.19$ & $3.28\pm0.30$ & $3.28\pm0.32$ & $0.07\pm0.37$ & N & $+$ \\

\hline\noalign{\smallskip}
\end{tabular}

\noindent
\end{table}
\section{Summary and Discussion}

We identified 50 PCF disappearance events using data obtained by SDO/AIA in 304~\AA\ and 211~\AA, and SDO/HMI magnetograms. Of these, 49 events showed clear eruption signatures, corresponding to approximately 98\% of the sample and are investigated in this study. Among these, 41 events are associated with identifiable CMEs. Our results show that: 

(a) As expected, 40 of the 41 associated CMEs were slow ($<800$ km s$^{-1}$);
(b) In our sample, PCF eruptions occurred both shortly before polar-field reversals and during relatively stable periods between reversals. No definitive conclusion can be drawn regarding a possible relationship between multiple polar-field reversals and PCF eruptions without further analysis of data from previous solar cycles;
(c) Rotational motions are observed in 24 of the 49 eruptive events, with identifiable rotation directions in 16 events. Among these events, nine allow further assessment of the magnetic configuration, and their rotation directions are consistent with the model predictions of \citet{Lynch2009}. This agreement supports the interpretation of these PCF eruptions as eruptions of flux-rope-like magnetic structures, consistent with both the modeling results of \citet{Chen2024} and the observational findings of \citet{Muhamad_2026};
(d) Ten of the 49 eruptive events were associated with TCHs, with magnetic flux ranging from $10^{20}$ to $10^{21}$ Mx, assuming a mean polar magnetic field of 10 G. Given the small number of TCH-associated events, further observations are needed to establish a general relationship between PCF eruptions and TCHs;
(e) Time--distance analysis shows that 22 events exhibit reliable two-phase kinematics, with onset heights mostly between 25 and 160 Mm and a mean onset velocity of about 2.8 km s$^{-1}$. No clear correlation is found between the low-coronal filament velocities and the associated CME speeds;
(f) Approximately 98\% (49 of 50) of the PCFs in our sample disappeared through eruption, while only one event was classified as a thermal disappearance. This predominance may be related to relatively weak magnetic confinement in polar regions. A sufficiently rapid decrease of the overlying coronal field with height, corresponding to a decay index exceeding the relevant critical threshold, could further facilitate PCF eruptions through torus instability \citep{Su2015}.

The previous study by \citet{Zhou_2020} established a correlation between the rotation direction of erupting filaments and their hemispheric location. Filaments in the northern hemisphere tend to exhibit counter-clockwise rotation, while those in the southern hemisphere generally prefer clockwise rotation during eruption. In our present study, the rotation directions observed in our limited sample of cases were consistent with this hemisphere-based model. However, given the small sample size, we cannot draw a firm conclusion that our results provide strong support for the direction-hemisphere relationship.
Beyond their rotation direction, the observed rotational motions, together with the inferred helicity signs, suggest that some PCFs are supported by magnetic flux ropes. The eruption of these structures provides a possible means of removing the filament plasma, linking the magnetic configuration to the observed disappearance.
In our sample, approximately 98\% of PCFs disappeared through eruption. One possible explanation is that the relatively weak overlying coronal field, with large enough decay index, provides less magnetic confinement, allowing the filaments to erupt more easily. Reconnection with the surrounding field may further reduce this confinement. For example, using stereoscopic observations from SDO and
Solar Orbiter, \citet{Muhamad_2026} studied the 2023 April 20 PCF eruption and proposed that breakout reconnection weakened the overlying confinement, allowing a pre-existing flux rope to rise and erupt.

Although this case illustrates how reconnection may facilitate a PCF eruption, it does not establish why eruptions dominate in our sample. We have not systematically examined the strength or geometry of the overlying field, both of which affect
magnetic confinement. With only one thermal disappearance event, we also cannot determine why thermal disappearance is uncommon in our sample.

While CMEs are closely linked to the solar polar magnetic field reversal, they are increasingly understood not as the primary cause, but rather as a consequence of this large-scale magnetic restructuring. Nonetheless, these eruptive events actively assist the reversal process. \citet{Low1997} proposed that CMEs act as a mechanism for the removal of old magnetic flux from polar regions, while \citet{Fisk2001} suggested that polar field reversals are driven by the reconnection of open field lines with closed loops—a process where CMEs can significantly help by facilitating and accelerating the magnetic reconnection. Further supporting this connection, \citet{Gopalswamy2003a} demonstrated a close relationship between high-latitude CMEs and the timing of polar field reversals during solar cycles 21 and 23. To evaluate this quantitatively, we can estimate the direct contribution of PCF eruptions to polar flux removal. The total magnetic flux in the polar areas is approximately $10^{22}$ Mx \citep{Benevolenskaya2004}. Because the magnetic flux removed by a typical eruption ($\Delta$Flux) is only on the order of $10^{19}$ to $10^{20}$ Mx, it would require approximately 100 eruptions to completely drive the polar reversal if eruptions were the primary cause. In our current sample, we directly observed three to four PCF eruption events (indicated by red dots with blue centers in Figure~\ref{Fig:butterfly}). Even assuming a similar number of events occurred on the far side of the solar disk (bringing the estimated total to approximately six to eight PCF eruptions prior to the observed reversal), this falls far short of the ~100 events needed to act as the primary driver. This discrepancy reinforces the view that these eruptions are a consequence of the ongoing polar reversal rather than its main cause, even though they actively facilitate the process by helping local reconnection and flux removal.

\textsc{Acknowledgments:} 
We wish to thank the referee for valuable comments that result in significant improvement of this paper.
This work is supported by NSF under grants AGS-2149748, 2300341 and by the NASA under grant  80NSSC20K1282 and cooperative agreement 80NSSC24M0174. YX acknowledge support from NSF RISE-2425602. JJ acknowledge support from NASA grants 80NSSC23K0406, 80NSSC21K1671, and 80NSSC21K0003. V.Y. acknowledges the support from NSF grants 2309939, 2408174 and 2401229, and NASA 80NSSC24K1914 grant. We thank the SDO/HMI and the SDO/AIA science teams for providing the full-disk magnetograms and EUV images.



\end{document}